\documentclass[%
 reprint, 
superscriptaddress,
 amsmath,amssymb,
 aps,prl
]{revtex4-1}
\usepackage{esvect}
\usepackage{graphicx,color}
\usepackage{dcolumn}
\usepackage{bm}
\usepackage[utf8]{inputenc}
\usepackage[T1]{fontenc}

\begin{document}

\preprint{APS/123-QED}
\title{Crystal orientation-dependent polarization state of high-order harmonics}%

\author{Yong Sing You}
\affiliation{Stanford PULSE Institute, SLAC National Accelerator Lab, Menlo Park, California}
\author{Jian Lu}
\affiliation{Stanford PULSE Institute, SLAC National Accelerator Lab, Menlo Park, California}
\author {Eric Cunningham}
\affiliation{Stanford PULSE Institute, SLAC National Accelerator Lab, Menlo Park, California}
\author{Christian Roedel} 
\affiliation{Institute of Optics and Quantum Electronics,University of Jena, Max-Wien-Platz 1,07743 Jena, Germany}
\author{Shambhu Ghimire}
\email{shambhu@slac.stanford.edu}
\affiliation{Stanford PULSE Institute, SLAC National Accelerator Lab, Menlo Park, California}

\date{\today}

\begin{abstract}
We analyze the crystal orientation-dependent polarization state of extreme ultraviolet (XUV) high-order harmonics from bulk magnesium oxide crystals subjected to intense linearly polarized laser fields. We find that only along high-symmetry directions in crystals high-order harmonics follow the polarization direction of the laser field. In general, the polarization direction of high-order harmonics deviates from that of the laser field, and the deviation amplitude depends on the crystal orientation, harmonic order and the strength of the laser field. We use a real-space electron trajectory model to understand the crystal orientation-dependent polarization state of XUV harmonics. The polarization analysis allows us to track the motion of strong-field-driven electron in conduction bands in two dimensions. These results have implications in all-optical probing of atomic-scale structure in real-space, electronic band-structure in momentum space, and in the possibility of generating attosecond pulses with time-dependent polarization in a compact setup.

\begin{description}

\item[PACS numbers]
42.65.Ky, 42.50.Hz, 72.20.Ht, 78.47.-p.

\end{description}
\end{abstract}

                             \maketitle

The process of high-order harmonic generation (HHG) in atomic targets has been widely studied and can be understood using a semiclassical three-step model comprising tunnel ionization, free-electron acceleration, and re-combination to the parent ion \cite{krause_high-order_1992,corkum_plasma_1993}. Because atomic targets exhibit spherical symmetries and the trajectories of accelerated electrons depend primarily on the laser parameters, high-order harmonics are polarized in the same direction as the driving laser pulse \cite{antoine_polarization_1997}. In crystals, the motion of strong-field driven electrons (and holes) is governed by the electronic band structure of the material so the usual strong-field approximation (SFA)\cite{keldysh_ionization_1965} is not appropriate without considering the role of the periodic potential. Consequently, depending on the band structure, the field driven electron trajectories may not necessarily follow laser polarization directions. These behavior could lead to a novel polarization states of solid-state HHG. 

In the solid-state HHG process \cite{vampa_theoretical_2014,wu_high-harmonic_2015}, a strong non-resonant laser field excites electrons from the valence band to the lowest conduction band, at the center of the Brillouin zone $k=0$ in direct band-gap materials such as ZnO. Then, the strong laser field $E(t)$ drives the electron (hole) in the conduction (valance) band towards the zone edge such that $k(t)=k(0)+\int E(\tau) d\tau$. Before dephasing occurs, the electron and hole can recombine at a particular crystal momentum, releasing the difference of their energy in the form of high-energy photons. At high-enough laser fields the electrons could be excited to a higher conduction band via Zenner tunneling \cite{zener_non-adiabatic_1932} and therefore the recombination to the hole in the valance band could yield even higher photon energy \cite{ndabashimiye_solid-state_2016,wu_multilevel_2016}. In addition, single-band nonlinear current could radiate depending on the non-parabolicity of the band \cite{ghimire_generation_2012,luu_extreme_2015}. Therefore, because of the band-motion of the electron the polarization of high-order harmonics may not necessarily be in the same direction as that of the laser field \cite{liu_high-harmonic_2017,luu_measurement_2018}. A careful analysis of polarization as a function of crystallographic orientation could be useful to track the  strong-field-driven motion of the electron (and hole), which makes solid-state HHG a novel tool to probe the electronic band structure in two dimensions. 
 
Recently, we used real-space electron trajectory model, which considers the strong-field-driven band motion of electron, to describe strongly anisotropic high harmonic response from wide band-gap, cubic MgO  crystals \cite{you_anisotropic_2017,you_probing_2018}. In this model, strong high-harmonic emission occurs when electron trajectories connect to the first and second nearest neighbor atoms in real-space while emission becomes weaker when the trajectories miss the atoms. The real-space electron trajectory model provides an additional possibility of reconstruction of the periodic potential with high sensitivity on valance charge density as optical excitations are naturally sensitive to the valance electrons. The electronic wave-function reconstruction requires the measurement of intensity, polarization, and spectral phase of high-order harmonics for various crystallographic orientations. 

Here, we investigate the crystal orientation dependent polarization state of XUV harmonics from MgO crystals pumped by strong near-infrared (NIR) laser fields. 
We find that the polarization of XUV harmonics is in the same direction as that of the laser pulse only when the laser field is aligned along high-symmetry directions of the crystal. As we rotate the crystal away from high-symmetry directions, the polarization of harmonics rotate in a non-trivial manner. The deviations depend on the crystal orientation, peak field strength of the laser pulse, as well as the harmonic order. To understand the origin of such a deviation, we extend the real-space electron trajectory model \cite{you_anisotropic_2017,you_probing_2018} and analyze collision angles, the angle between the direction of the laser field and the direction of group velocity at the time of collisions to the nearest neighbor atoms. Because the group velocity in two dimensional momentum space is defined by the anisotropic band-structure its instantaneous value can deviate significantly from the direction of the laser field, causing a substantial rotation of the harmonic's polarization.

\begin{figure}
\begin{center}
\includegraphics[width=3.375in, trim =0cm 0cm 0cm 0cm, clip = true]{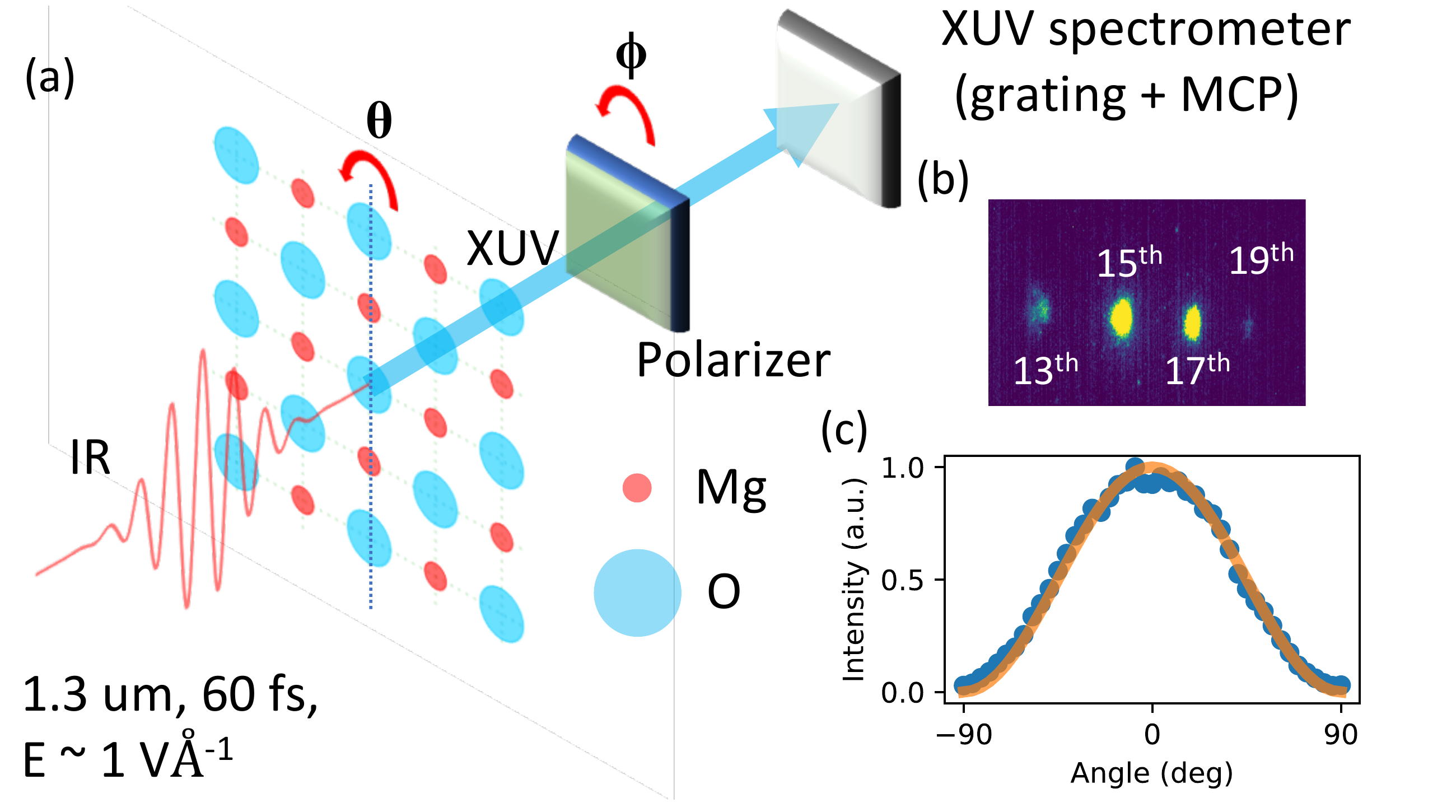}
\caption{(a) Experimental setup for analyzing the polarization state of extreme ultraviolet (XUV) high-harmonics. XUV harmonics are produced in wide band-gap MgO crystals pumped by 1.3 $\mu$m, 60 $\mathrm{fs}$ laser pulses of estimated field strengths up to $\sim$ 1$V/\mbox{\AA}$ without damage. $\theta$ and $\phi$ are crystal and polarizer angles respectively. The polarizer consists of three gold-coated mirrors arranged together such that the entire assembly can be rotated without changing the XUV beam direction (see details in Hahn et al.\cite{hahn_broadband_2015}). The raw spectrum is shown at (b). (c) Measurement of the polarization state of 17$^\mathrm{th}$ order harmonic.The measured data (dots) fits well with a $cos^2\phi$ fitting curve (solid line) such that the maxima is at $\phi$=0 and the minima is at $\phi$= +/-90 degrees, indicating a linear polarization along 0 degree.}
\label{f:figure1}
\end{center}
\end{figure}

\begin{figure}
\begin{center}
\includegraphics[width=3.375in, trim =0cm 0cm 0cm 0cm, clip = true]{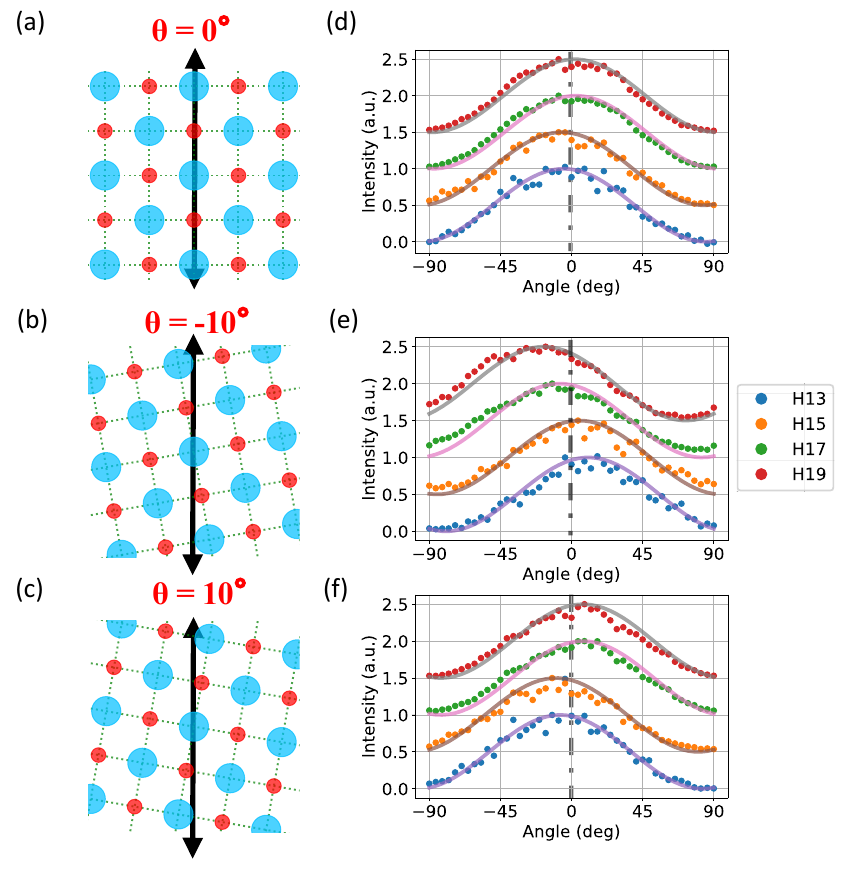}
\caption{(a)-(c) Rotation of MgO crystal with respect to the laser polarization direction. (d)-(f) Measurement of polarization state of high-order harmonics for different crystal angles. Experimental data and their fits for different harmonic orders are offset vertically for clarity. In (d) all harmonics follow laser polarization direction. In (e) and (f) harmonics show strong deviations upto 20 degrees.}
\label{f:figure2}
\end{center}
\end{figure}

The experimental setup is shown in Figure \ref{f:figure1}. Laser pulses of 60 femtosecond ($\mathrm{fs}$) duration and a central wavelength of 1.3 $\mu$m are focused into a 001-cut, 200 $\mu$m thick MgO crystal at normal incidence. The laser pulses are focused with a 40 cm lens to achieve a peak field strength as high as 1 V/\mbox{\AA} without damaging the sample. High harmonics emitted in the transmission geometry are measured using an imaging XUV spectrometer, consisting of a flat-field variable groove density grating and a micro-channel plate detector. Figure \ref{f:figure1}(b) shows a representative measured spectrum with a high-energy cutoff around $19^\mathrm{th}$ order harmonic. We use an XUV polarizer based on the Fresnel reflections of three gold mirrors in grazing incidence \cite{hahn_broadband_2015}. We keep the laser polarization fixed and rotate the crystal and the XUV polarizer as needed. First, we rotate the crystal such that the cubic (Mg-O bonding direction) direction of MgO crystal is along the laser field. We measure the polarization of high-harmonics (HH-polarization) by rotating the polarizer. Figure \ref{f:figure1}(c) shows the polarization measurement of the $17^\mathrm{th}$ order harmonic. The measured data fits well with a $cos^2\phi$ fitting curve, where $\phi$ is the angle between the polarization direction of the laser field and optic axis of the XUV polarizer. The extinction ratio is $\sim$ 1/50. Figure \ref{f:figure2}(a) shows the crystallographic orientation with respect to the laser field (orientation angle $\theta$), and (d) shows the polarization state for all detected harmonics, from $13^\mathrm{th}$ to $19^\mathrm{th}$ order. It is seen that the maxima ($\phi = 0^\circ$) and minima ($\phi = 90^\circ$) for all harmonics line up well, suggesting that polarization of all harmonics follow the laser field polarization along high-symmetry direction. 

Next, we rotate the crystal such that the laser field direction is away from the high-symmetry direction of the crystal, and we measure the polarization state of high harmonics. Figure \ref{f:figure2}(b) shows a case when the crystal is rotated 10 degrees to the left, i.e. $\theta=-10^{\circ}$, and in this case we find that the maxima angles on polarizer scans (e) are shifted from the direction of the laser field ($\phi=0^\circ$, $\theta=0^\circ$) upto about 20 degrees. $13^\mathrm{th}$ and $15^\mathrm{th}$ HH are polarized away from the Mg-O bonding while $17^\mathrm{th}$ and $19^\mathrm{th}$ HH are polarized closer to the Mg-O bonding direction. Similarly, we rotate the crystal to the right by 10 degrees, i.e. $\theta=10^{\circ}$, as shown in Figure \ref{f:figure2}(c) and in this case we observe a reverse trend (left and right) in the polarization shift, as $13^\mathrm{th}$ and $15^\mathrm{th}$ order shift further from Mg-O bonding direction and $17^\mathrm{th}$ and $19^\mathrm{th}$ order shift closer to the Mg-O bonding direction. We note that the extinction ratio in the polarization scans did not change when we rotated the crystal. It means that, within the measurement error-bars, the harmonics are always linearly polarized. Also, we verified that the driving laser pulse does not become elliptically polarized after sample.

\begin{figure}
\begin{center}
\includegraphics[width=3.375in, trim =0cm 0cm 0cm 0cm, clip = true]{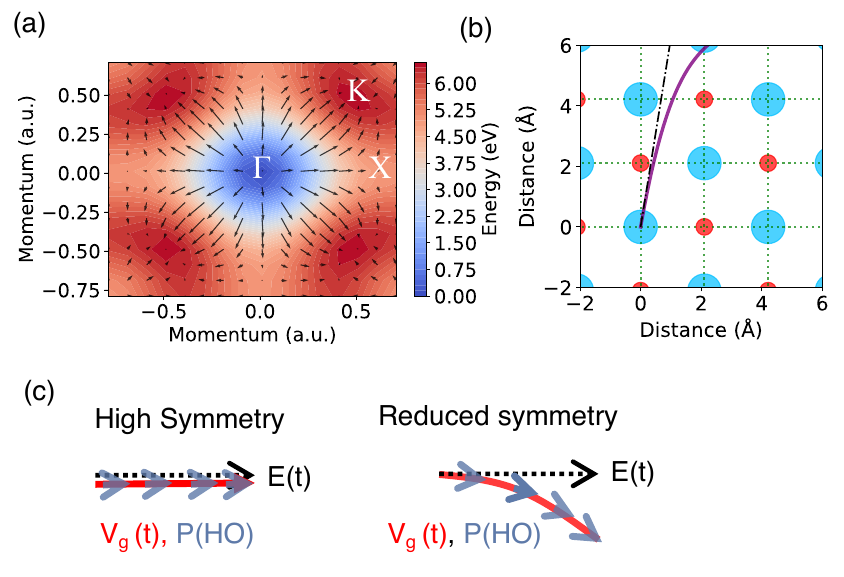}
\caption{(a) Conduction band (CB1) of MgO considering simple tight-binding model. Arrows represent the amplitude and direction of group velocity   $\vec{v_g}(t)$ (b) Classical electron trajectory (solid line) under strong laser field (dotted line).The direction of the electron trajectory deviates from the laser polarization direction when the laser field is not along high-symmetry directions. (c) The polarization direction high-harmonics $P(HO)$ is determined by the direction of  $\vec{v_g}(t)$ at the instant of collision to atomic sites. In high-symmetry directions, the HH polarization follow laser polarization (left); at reduced symmetry directions, they deviate from the laser polarization (right).}
\label{f:figure3}
\end{center}
\end{figure}

In order to understand the crystal orientation dependence of the harmonics polarization, we use the real-space electron trajectory model \cite{you_anisotropic_2017}. Previously, we used this model to describe how the total yield of high-harmonics changes when the crystal is rotated with respect to the laser polarization. The model consists of three steps, in analogy to the semi-classical three-step re-collision model. The model also considers collisions to the nearest neighbor atoms and the electron dynamics repeat every unit cell because of delocalized electron wave-packets. In the first step, we assume that the electron tunnels from the highest valance band to the lowest conduction band near the peak of the laser field $E(t)=E_0cos(\omega t)$. Second, the electron in the conduction band accelerates as it acquires momentum and energy from the laser field. The equation of motion of the electron is given by $\frac{d\vec{k}}{dt}=-\vec{E}(t)$ (in atomic units). The electron trajectories, $\vec{r}$, are obtained by numerically solving $\frac{d\vec{r}}{dt}=\frac{\partial\epsilon(\vec{k})}{\partial\vec{k}}$. In the third step, the electron collides coherently with the atomic cores, releasing energy in the form of high-energy photons. In this step, the direction of the harmonics's polarization is given by the direction of the instantaneous group velocity $\vv{v_g}(t)=\frac{d\vec{r}}{dt}$ at the time of collision with respect to the laser field. Figure \ref{f:figure3} shows how $\vv{v_g}(t)$ may or may not be along the direction of the laser field depending on if the laser field is along a direction of high symmetry. Figure \ref{f:figure3}(a) shows the first conduction band of a 001-cut MgO crystal in two-dimensions along with the distribution of $\vv{v_g}(t)$ such that the direction, and length of arrow represent the direction and amplitude of $\vv{v_g}(t)$ at certain instant of time $t$. Figure \ref{f:figure3}(b) shows a representative electron trajectory in the real-space when the laser field is slightly off from the cubic direction. In this case, it can be seen that the electron trajectory (solid line) deviates from the direction of the laser field (dashed line). This deviation in group velocity, as illustrated schematically in Figure \ref{f:figure3}(c), is the origin of the deviation of the harmonics' polarization with respect to the laser field. 

\begin{figure}
\begin{center}
\includegraphics[width=3.375in, trim =0cm 0cm 0cm 0cm, clip = true]{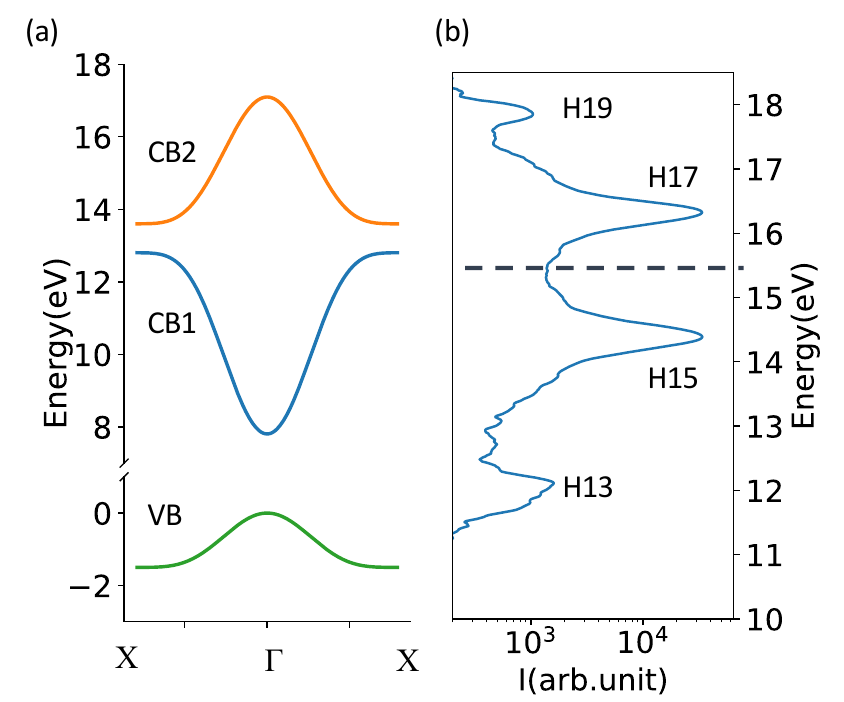}
\caption{(a) 
A portion of electronic band structure of MgO crystal showing important bands; one of the fully occupied valence bands (VB), which are degenerate at $\Gamma$ point, lowest conduction band (CB1) and a higher-lying dipole allowed conduction band (CB2). (b) A portion of high harmonic spectrum, limited by collection angle of the spectrometer (lower photon energy side). The dash line indicates the approximate location of the band-gap at $\boldmath{X}$ point VB-CB1. 13$^\mathrm{th}$ and 15$^\mathrm{th}$ harmonics are below this limit and 17$^\mathrm{th}$ and 19$^\mathrm{th}$ harmonics are above this limit.}
\label{f:figure4}
\end{center}
\end{figure}

We begin our analysis by assigning the photon energies of the harmonics to relevant bands in the electronic band structure following previous work \cite{wu_orientation_2017,you_laser_2017}. According to the inter-band model, the maximum photon photon energy of harmonics should be limited by the maximum bandgap between participating electronic bands. Figure \ref{f:figure4}(a) shows a portion of the band structure of MgO, revealing three important bands: one of the highest lying valance bands (VB), the lowest conduction band (CB1), and a higher-lying conduction band (CB2). A representative harmonic spectrum is plotted on the side in \ref{f:figure4}(b). It is seen that the photon energy of the $13^\mathrm{th}$ and $15^\mathrm{th}$ order harmonics can be accommodated by the band pair VB-CB1, which has a maximum band-gap of $\sim$15 eV at the zone edge. The photon energy of the $17^\mathrm{th}$ and $19^\mathrm{th}$ order harmonic are higher than this so their origin is assigned to the pair VB-CB2. We note that the measured spectrum in its lower photon energy end is limited by the collection angle of the spectrometer.

\begin{figure}
\begin{center}
\includegraphics[width=3.375in, trim =0cm 0cm 0cm 0cm, clip = true]{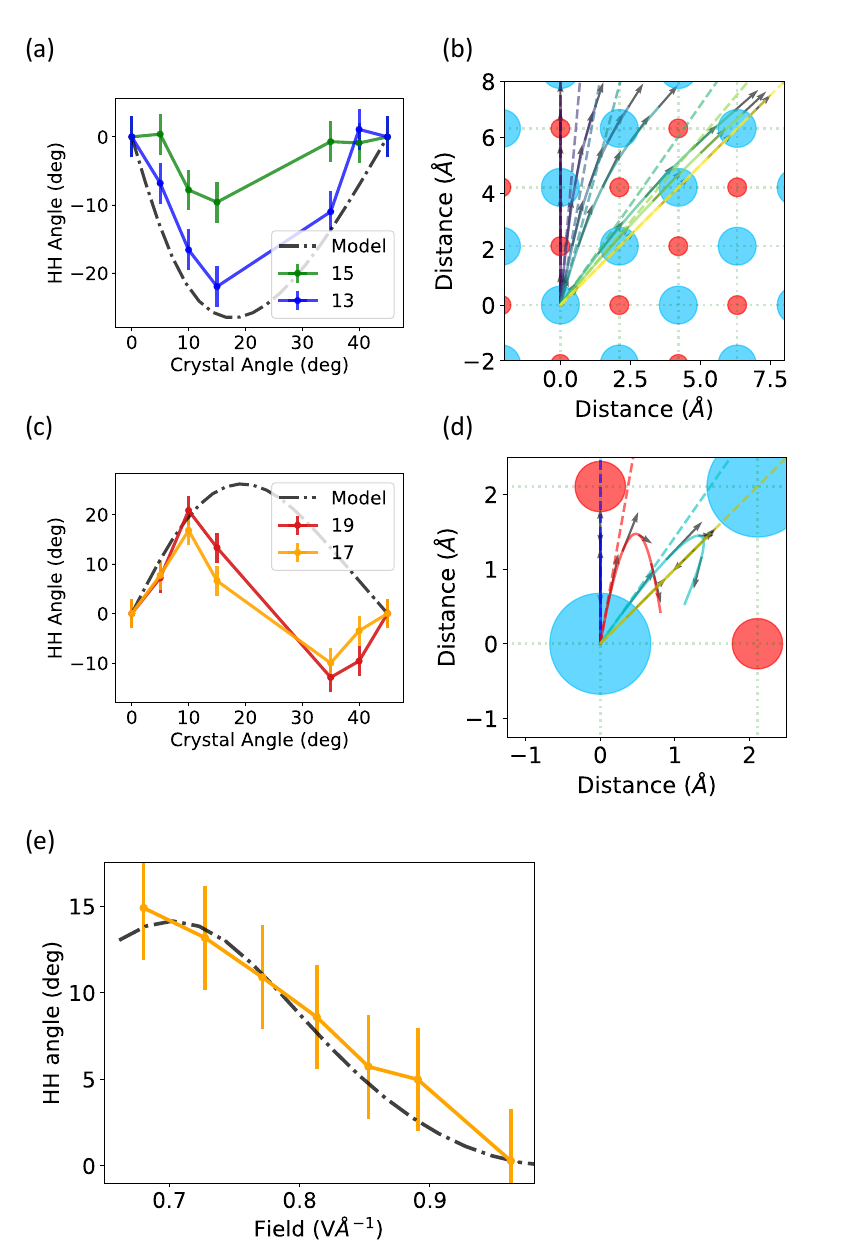}
\caption{(a) Measured polarization rotation of 13$^\mathrm{th}$ and 15 $^\mathrm{th}$ order harmonics as a function of crystal  angles. Calculated rotation angles are also shown (dash-dotted line) for a qualitative comparison. (b) Electron trajectories using band CB1. (c) Polarization rotation of $17^\mathrm{th}$  and $19^\mathrm{th}$ order harmonics versus crystal angles. Calculated polarization rotation angles are also shown for a qualitative comparison. (d) Electron trajectories  using band  CB2. (e) Laser field strength dependence of   polarization rotation, experimental (solid line) and simulation results (dashed and dotted black line).}
\label{f:figure5}
\end{center}
\end{figure}

In experiments, we rotate the crystal in small steps and analyze polarization of high-harmonics. First, we show the results of $13^\mathrm{th}$ and $15^\mathrm{th}$ order harmonics in Figure \ref{f:figure5}(a),(b). Their origin is assigned to VB-CB1. When the crystal angle is $\theta = 0^\circ$, the cubic direction (Mg-O bonding direction) aligns to the laser polarization direction. We rotate the crystal to the right (positive angles) such that crystal angles are 5, 10, and 15 degrees, and we observe that harmonics rotate to the left (negative angles) by 7, 16, and 22 degrees respectively. The field strength of the laser pulse is 0.7 $V/\mbox{\AA}$. We observe that the polarization deviation first increases with the crystal rotation angle and then it decreases fairly reversibly going back to the direction of the laser field at 45 degree crystal angle (Mg-Mg/O-O directions), which is another high symmetry direction. In Figure \ref{f:figure5} (b), we plot the real-space electron trajectories considering various crystal angles. It is seen that along high-symmetry directions (blue and magenta lines), the electron group velocity follows the laser field direction. However, at other angles (green, red and cyan lines), the electron group velocity deviates away from the laser field; moreover, this set of harmonics deviates towards the O-O direction. The corresponding calculated polarization directions for various crystal angles are shown in Figure \ref{f:figure5}(a), which qualitatively reproduce experimental results. 

Second, we discuss the behavior of $17^\mathrm{th}$ and $19^\mathrm{th}$ order harmonics, whose origin is assigned to the band-pair VB-CB2. As seen in Figure \ref{f:figure5}(c), at crystal angles $\theta = 0^\circ$ and $\theta = 45^\circ$ these harmonics are also along the laser polarization direction, rotating as we rotate the crystal away from these high-symmetry directions. However, the rotation of these harmonics is in the opposite direction compared to the first set of harmonics, rotating instead towards the cubic direction of the crystal. Figure \ref{f:figure5}(d) shows calculated real-space electron trajectories considering the band motion on CB2, for various crystal angles. Again, it is seen that along high-symmetry directions (blue and yellow lines), the electron group velocity follows the laser field direction. However, at other angles (red and cyan lines), the trajectories deviate strongly from the direction of laser field. The rich oscillatory behavior of the electron arises from its localization in real-space, as the electron in momentum space experiences periodic Bragg-like diffractions at Brillouin zone edges \cite{mucke_isolated_2011,ghimire_generation_2012,luu_extreme_2015}. We note that here the tunneling to CB2 occurs predominantly at the zone edge ($\boldmath{X}$) because of the smaller band-gap, which is in contrast to the VB-to-CB1 tunneling since the band-gap there is smallest at the zone center ($\Gamma$ point). The calculated polarization directions for various crystal angles are plotted in Figure \ref{f:figure5}(c), which qualitatively reproduce experimental results. We also note that the amplitude of the polarization deviation on all harmonics strongly depends on the maximum strength of the laser field. 

Finally, to understand how the polarization shift depends on the laser field strength, we perform a systematic study of the shift in polarization angle as a function of peak field strength of the pump laser. In particular, we examine the behavior of $17^\mathrm{th}$ order harmonic for a crystal angle of $\theta=10^\circ$. We tune the laser field from 0.68 to 0.96 $V/\mbox{\AA}$. In Figure \ref{f:figure5}(e), we show experimental results (dots) and simulation results (dashed line). A reasonable agreement between simulation results and experimental observation can be seen in that the amplitude of rotation decreases with increase in the strength of the laser field.

These theoretical and experimental results show that when the laser field is not along high-symmetry directions in the crystal, the driven electron experiences unequal accelerations in two-dimensional momentum space, and therefore the instantaneous group velocity does not follow the direction of the laser field. Consequently, the polarization of high-harmonics deviates from the polarization of the laser field. Essentially, band dispersions play defining roles in polarization deviations of corresponding high-order harmonics. In MgO, two conduction bands have opposite dispersion, and therefore two set of harmonics (VB-CB1 and VB-CB2) rotate in opposite directions. A careful analysis of harmonic order dependence of the polarization deviation allows us to retrieve instantaneous group velocity of the strong-field-driven electron in conduction bands in two dimensions. This is analogous to molecular HHG, where the study of anisotropic high-harmonic responses \cite{itatani_tomographic_2004} and the analysis of polarization state of high-harmonics \cite{levesque_polarization_2007,zhou_elliptically_2009} reveal important information about the participating electronic orbital of the molecule.  However, in solids underlying microscopic processes are complex in part because of the possibility of coupled inter- and intra-band processes \cite{tancogne-dejean_ellipticity_2017,hohenleutner_real-time_2015} and poorly understood dephasing mechanisms \cite{vampa_linking_2015,vampa_all-optical_2015}. In MgO crystals, the observed harmonic order dependent polarization shift indicates strongly that inter-band processes dominate. Our findings presented here are important steps towards using anisotropic high-harmonic response of crystals to reveal atomic-scale structure in the real-space and electronic band-structure in the momentum space in two dimensions. Also, high-harmonic polarimetry could be a powerful approach to investigate novel solid-state processes that are difficult to access otherwise. Recently, Luu et al \cite{luu_measurement_2018} used high-harmonic polarimetry to identify the role of Berry's curvature in high-harmonic generation from SiO$_2$. Other implications of our results include ultrafast switching of XUV pulse, for example, by choosing proper crystal orientations the polarization of $15^\mathrm{th}$ order harmonic can be rotated against that of $13^\mathrm{th}$ order harmonic by about 10 degrees. Our results show that solid-state HHG could support attosecond pulse with time-dependent polarization, which could be useful to probe charge migration \cite{kuleff_core_2016}. 

This work is funded by the U.S. Department of Energy, Office of Science, Basic Energy Sciences, Chemical Sciences, Geosciences, and Biosciences Division through the Early Career Research Program. C.R. acknowledges support from the Volkswagen Foundation.

%

\end{document}